# Thermodynamic Screening of Metal-Substituted MOFs for Carbon Capture

Hyun Seung Koh,[a] Malay Kumar Rana,[a] Jinhyung Hwang,[a] and Donald J. Siegel[*,a]

Metal-organic frameworks (MOFs) have emerged as promising materials for carbon capture applications due to their high $CO_2$ capacities and tunable properties. Amongst the many possible MOFs, metal-substituted compounds based on M-DOBDC and M-HKUST-1 have demonstrated amongst the highest $CO_2$ capacities at the low pressures typical of flue gasses. Here we explore the possibility for additional performance tuning of these compounds by computationally screening 36 metal-substituted variants (M = Be, Mg, Ca, Sr, Sc, Ti, V, Cr, Mn, Fe, Co, Ni, Cu, Zn, Mo, W, Sn, and Pb) with respect to their $CO_2$ adsorption enthalpy, $\Delta H^{T=300K}$. Supercell calculations based on van der Waals density functional theory (vdW-DF) yield enthalpies in good agreement with experimental measurements, out-performing semi-empirical (DFT-D2) and conventional (LDA & GGA) functionals. Our screening identifies 13 compounds having $\Delta H$ values within the targeted thermodynamic window $-40 \leq \Delta H \leq -75$ kJ/mol: 8 are based on M-DODBC (M=Mg, Ca, Sr, Sc, Ti, V, Mo, and W), and 5 on M-HKUST-1 (M= Be, Mg, Ca, Sr and Sc). Variations in the electronic structure and the geometry of the structural building unit are examined and used to rationalize trends in $CO_2$ affinity. In particular, the partial charge on the coordinatively unsaturated metal sites is found to correlate with $\Delta H$, suggesting that this property may be used as a simple performance descriptor. The ability to rapidly distinguish promising MOFs from those that are "thermodynamic dead-ends" will be helpful in guiding synthesis efforts towards promising compounds.

## Introduction

Anthropogenic carbon dioxide ($CO_2$) emissions from combustion of fossil fuels is the largest contributor to global climate change.[1] Moreover, continued population growth and economic development are expected to increase fossil fuel consumption.[2,3] Consequently, increasing emphasis has been placed on the development of carbon capture and sequestration (CCS) technologies aimed at reducing the emissions of carbon-intensive point sources such as coal-fired power plants.[4,5]

Given their potential for facile regeneration, physisorptive materials present an attractive alternative to current monoethanolamine (MEA)-based approaches that use chemisorptive interactions to capture $CO_2$.[6] In particular, metal-organic frameworks (MOFs)[7-11] have emerged as promising solid sorbents for $CO_2$ capture under conditions relevant for flue gas applications.[12-16] MOFs are micro-porous crystalline materials constructed from metal ions or polynuclear metal clusters assembled in a periodic fashion through coordination to organic ligands (referred to as linkers) resulting in an extended host structure. The building block nature of MOFs allows for the synthesis of a wide variety of crystal structures and a remarkable tunability in their properties.[17] Although MOFs are a relatively new class of materials, they have already exceeded the records set by zeolites in the regime critical for flue gas capture, i.e., low partial pressures of $CO_2$.[12]

While many MOFs have been synthesized,[18] (and many thousands more have been hypothesized[19],) the sub-set of compounds exhibiting coordinatively unsaturated metal sites (CUS), have demonstrated amongst the highest $CO_2$ capacities.[20] CUS have been identified as the primary adsorption site for $CO_2$,[21,22] and enable $CO_2$ uptake at low pressures[23] and moderate pressures due to their strong interaction with adsorbates. For example, Mg-DOBDC is a MOF composed of unsaturated $Mg^{2+}$ ions in square-pyramidal coordination arranged in linear, infinite chains linked by 2,5-dioxido-1,4-benzene dicarboxylate (DOBDC). This material surpasses all other sorbents in its ability to concentrate $CO_2$ from dilute streams: at 296 K and 0.1 atm Mg/DOBDC adsorbs 5.4 mol $CO_2$/kg (~200 kg/m$^3$ on a volumetric basis).[12]

In addition to the possibility for high $CO_2$ capacities, recent studies have shown that metal substitution can be used to tune the performance of these compounds.[12,24] For example, Caskey et al.[12] demonstrated that isostructural substitutions of Mg, Ni, and Co for Zn in Zn-DOBDC could dramatically alter $CO_2$ affinity;[25] more recent work has reported the synthesis of Fe-,[26] Mn-,[27] and Cu-substituted[28] variants. HKUST-1 is another noteworthy example of a CUS-containing MOF that has demonstrated both high $CO_2$ capacity and potential for metal substitution.[29] In this MOF the metal cluster adopts a paddle wheel geometry with a square-planar coordination of the CUS metal.[21] In this case Cr[30], Ni[31], Zn[32], Ru[33], and Mo-substituted[34] variants of the original Cu-based compound have been synthesized.[35] As the CUS are the primary binding site for $CO_2$, the ability to substitute different CUS metals implies the ability to tune the adsorption enthalpy, $\Delta H$, and consequently several properties important for $CO_2$ capture: $\Delta H$ has been linked to the capacity,[20,23,36,37] selectivity,[23] and regeneration efficiency of a given MOF.[6,38] Regarding regeneration efficiency, it has been suggested that optimal physisorbents for CCS should exhibit adsorption enthalpies within a range of $-40$ to $-75$ kJ/mol.[38]

In light of the importance of ΔH, the ability to accurately predict its magnitude across a range of MOFs would be helpful in screening for optimal CCS compounds. Recently, van der Waals density functional methods, vdW-DFT[39], have demonstrated promising accuracy under various environments[39-47] to account for long-ranged dispersion interactions within conventional DFT at moderate computational cost. For example, in our previous study[48] several van der Waals density functionals (vdW-DFs) were benchmarked against experimental adsorption enthalpies in 4 CUS-containing MOFs: Mg-[12], Ni-[49], Co-DOBDC[50] and Cu-HKUST-1.[29] Comparisons were made between conventional LDA[51] and GGA[52] functionals (with no dispersion interaction), the semi-empirical DFT-D2[53], and vdW-DF's with five distinct GGA-based exchange functionals: revPBE[39], optB86[54, 55], optB88[54], optPBE[54, 55], and rPW86.[56] The calculations revealed that the revPBE-vdW functional produced very good agreement with the average experimental enthalpies, with an error of ~2 kJ/mol, suggesting that thermodynamic screening based on vdW-DF's is computationally feasible, even for MOFs with large unit cells (e.g., HKUST-1).

Towards the goal of identifying thermodynamically-optimal MOFs for CCS applications,[57] in this study we computationally screen 36 metal-substituted variants of M-DOBDC and M-HKUST-1 (M = Be, Mg, Ca, Sr, Sc, Ti, V, Cr, Mn, Fe, Co, Ni, Cu, Zn, Mo, W, Sn, and Pb) with respect to their $CO_2$ adsorption enthalpy, $\Delta H^{T=300K}$. Supercell calculations based on the revPBE-vdW functional yield enthalpies in good agreement with experimental measurements, out-performing semi-empirical (DFT-D2) and conventional (LDA & GGA) functionals. Our screening identifies 13 compounds having $\Delta H$ values within the targeted thermodynamic window $-40 \le \Delta H \le -75$ kJ/mol: 8 are based on M-DODBC (M=Mg, Ca, Sr, Sc, Ti, V, Mo, and W), and 5 on M-HKUST-1 (M= Be, Mg, Ca, Sr and Sc). Variations in the electronic structure and the geometry of the structural building unit are examined and used to rationalize trends in $CO_2$ affinity. In particular, the partial charge on the coordinatively unsaturated metal sites correlates with $\Delta H$, suggesting that this property may be used as a simple performance descriptor. The ability to rapidly distinguish promising MOFs from those that are "thermodynamic dead-ends" will be helpful in guiding synthesis efforts towards promising compounds.[38]

## Methods

Thermodynamic screening of $CO_2$ adsorption enthalpies across metal-substituted variants of M-DOBDC and M-HKUST-1 was performed using van der Waals-augmented density functional theory (DFT[58], VASP[59, 60] code). Crystal structures for empty[29, 49] and $CO_2$-containing[21, 61] Ni-DOBDC and Cu-HKUST-1 were adopted from diffraction experiments and were used as initial models for metal-substituted versions in which the metal component (M) was selected from elements which have the potential to exhibit a +2 oxidation state. These include: four alkaline earths: Be, Mg, Ca, Sr; 11 transition metals: Sc, Ti, V, Cr, Mn, Fe, Co, Ni, Cu, Zn, Mo, W, and two group-14 metals: Sn, Pb.

The computational cells used for M-DOBDC and M-HKUST-1 contain, respectively, 54 and 156 atoms; 6 and 12 $CO_2$ molecules were added to these supercells to represent the adsorbed state, corresponding to a coverage of one $CO_2$ per CUS. For M-DOBDC, the symmetry for both empty and $CO_2$-containing supercells adopts the $R\bar{3}$ space group, as found in experiments[49]. In the case of M-HKUST-1, $CO_2$ adsorption at CUS sites can occur in one of four symmetry-equivalent positions, (each having a $CO_2$ occupancy of 25%,) which differ by a 90° rotation about an axis connecting the two metal sites within an SBU[21]. To account for the fractional occupancy, M-HKUST-1 supercells were constructed such that the $CO_2$ molecules occupy one of the 4 possible adsorption sites in a quasi-random fashion: the two $CO_2$ molecules adsorbed on opposite sides of an SBU were positioned in a *trans* configuration (see Fig. S1 in the Supporting Information) to maximize the $CO_2$-$CO_2$ separation, and this *trans* configuration was varied from SBU to SBU throughout the cell. Consequently, supercells containing adsorbed $CO_2$ have a slightly lower symmetry [$Fm\bar{3}$ (#202)] than those without [$Fm\bar{3}m$ (#225)].

Static binding energies for $CO_2$ at zero Kelvin ($\Delta E$) were calculated using two dispersion-corrected versions of DFT: vdW-DF1[39], and the semi-empirical DFT-D2[53] method. For comparison, energies calculated using the "conventional" (i.e., without dispersion corrections) Ceperley-Alder LDA[51] and PBE-GGA[52] functionals are also reported. In all cases geometries were relaxed to a force tolerance of 0.01 eV/Å. As previously demonstrated[48], the revPBE-vdW functional in vdW-DF1[39] yields excellent agreement with experimental $CO_2$ adsorption enthalpies in prototypical CUS-MOFs, outperforming other dispersion-corrected functionals[54-56] and DFT-D2[53]. Room temperature (T = 300 K) adsorption enthalpies ($\Delta H$) were computed by adding zero point energy (ZPE) and thermal contributions ($\Delta TC$) to the static binding energies ($\Delta E$). All calculations were spin-polarized and performed with a plane-wave energy cut-off of 500 eV; k-point sampling was performed at the $\Gamma$-point, and yielded adsorption energies converged to less than 1 kJ/mol $CO_2$. (For density of states calculations a denser k-grid of 2x2x2 was used.) The interactions between core and valence electrons were described by the projector-augmented-wave (PAW) method[60] in which the semi-core electron states are treated as valence states. Atomic charges were evaluated using the REPEAT method.[62] Additional details regarding the calculation methods can be found in Ref. 46.

Static binding energies ($\Delta E$) at 0 K were calculated using the following expression:

$$\Delta E = \frac{1}{n}(E_{MOF+CO_2} - E_{MOF} - nE_{CO_2}),$$

where, $E_x$ refers, respectively, to the total energies of the MOF+$CO_2$ complex, the isolated MOF, and an isolated $CO_2$ molecule. $n$ is the total number of adsorbed $CO_2$ molecules. Adsorption enthalpies at T = 300 K were then calculated as follows:

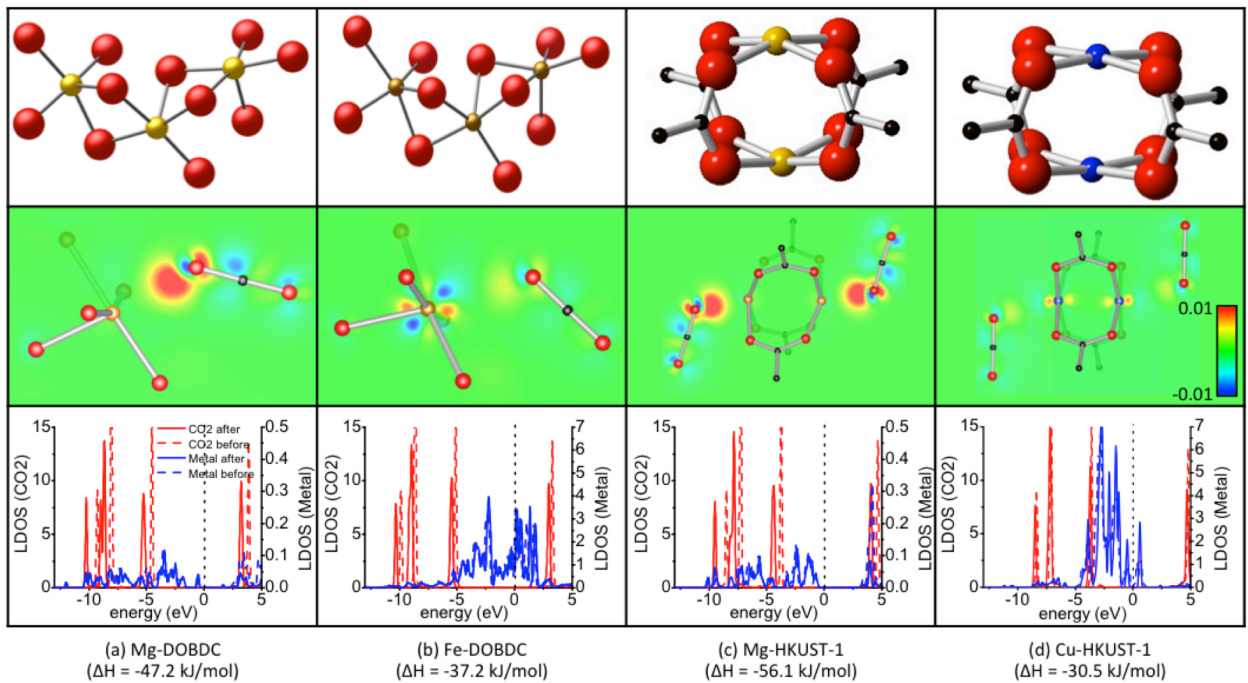

**Figure 1.** Local geometry (top row), charge density difference (middle), and local DOS (LDOS, bottom) for four representative MOFs in the vicinity of the SBU. From left to right: Mg-DOBDC, Fe-DOBDC, Mg-HKUST-1, and Cu-HKUST-1. C = black; O = red; Mg = yellow; Fe = dark yellow; Cu = blue. For clarity, in the case of DOBDC only a portion of the infinite SBU is shown in the top row, while for the charge density difference (units of electrons/Å$^3$) only the metal and its nearest neighbors are shown. For the LDOS plot solid lines refer to the adsorbed state, and dashed lines refer to isolated CO$_2$ and MOF.

$$\Delta H = \Delta E + \Delta E_{ZPE} + \Delta E_{TE},$$

where the total enthalpy of a system is given by

$$H = E0 + E_{ZPE} + E_{TE}.$$

In the above, E0 is the 0 K static total energy, $E_{ZPE} = \Sigma_i \frac{\hbar \omega_i}{2}$ and $E_{TE} = \Sigma_i \frac{\hbar \omega_i}{\exp\left(\frac{\hbar \omega_i}{k_B T}\right) - 1}$ are the zero point energy and vibrational contributions. $k_B$ is the Boltzmann constant and $w_i$ are the normal mode vibrational frequencies. In case of free CO$_2$, an additional $7/2\, k_B T$ is added to $E_{TE}$ to account for translational, rotational, and PV degrees of freedom.

## Results and Discussion

### Structure

While isostructural variants of DOBDC and HKUST-1 have been reported for several metal compositions[12, 24-27, 29-34, 49, 50], it is unclear if variants beyond those currently known are possible. Here we examine the relaxed structure of the SBU as a qualitative indicator of whether a given isostructural variant is plausible. Figures S2 and S3 depict the local coordination of the CUS in M-DOBDC and HKUST-1 as a function of the substituted metal, based on relaxations with the revPBE-vdW functional. (Optimized cell geometry data appears in Table S1; metal-oxygen bonding distances are given in Tables S2 and S3.) Each structure is classified according to a "red/yellow/green" color scheme based on the extent to which the relaxed structure resembles its respective prototype structure, Ni-DOBDC or Cu-HKUST-1. Structures labeled "green" exhibit geometries that are very similar to the prototype, indicating that isomorphism may be possible. On the other hand, "red" structures exhibit large structural distortions, such as a change in the coordination number of the CUS. We expect that these compounds are less likely to exhibit isomorphism. Finally, "yellow" compounds fall between these extremes, and refer to systems in which there are moderate structural deviations (e.g., changes in bond length) from the prototype.

In the case of M-DOBDC, 13 of the 18 candidate structures (excluding Be, Cr, Cu, Sn and Pb) exhibit only minor changes to the SBU geometry, and therefore fall within the green category. These compounds maintain the square-pyramidal coordination of the CUS to its nearest-neighbor oxygens. Moreover, bond lengths for the five M-O bonds follow trends similar to those observed in the Ni-DOBDC prototype (Table S2). The variant containing substituted Pb falls within the yellow category as it exhibits a slightly distorted structure with a much wider range of M-O distances of 2.328 – 2.696 Å (Table S2). Finally, structures containing Be, Cr, Cu and Sn fall within the red category because their geometries contain CUS with a coordination number of four. In particular, Be adopts a tetrahedral coordination and becomes "buried" inside the MOF structure. This behavior appears to arise from the small ionic radius of the Be$^{+2}$ ion. In this geometry Be is no longer accessible to guest molecules, explaining the relatively low adsorption enthalpy observed in this compound (see below).

| Metal | M-O (Å) | Metal Charge | ΔH (kJ/mol) | | |
|---|---|---|---|---|---|
| | | | DFT-D2 | rPBE-vdW | Experiments |
| M-DOBDC | | | | | |
| Be | 4.049 | 1.248 | -15.6 | -31.8 | |
| Mg | 2.392 | 1.556 | -38.5 | -47.2 | -44.2±4.6[12, 63-67] |
| Ca | 2.623 | 1.487 | -36.1 | -46.3 | |
| Sr | 2.842 | 1.460 | -30.4 | -44.6 | |
| Sc | 2.406 | 1.553 | -41.8 | -51.5 | |
| Ti | 2.394 | 1.738 | -47.9 | -52.7 | |
| V | 2.276 | 1.520 | -52.2 | -53.5 | |
| Cr | 3.286 | 1.082 | -17.4 | -32.9 | |
| Mn | 2.695 | 1.191 | -28.9 | -37.3 | |
| Fe | 2.717 | 1.309 | -23.6 | -32.4 | |
| Co | 2.812 | 1.099 | -28.9 | -37.2 | -35.7±1.9[12, 66] |
| Ni | 2.617 | 1.173 | -29.3 | -37.7 | -39.6±1.5[12, 61, 66] |
| Cu | 3.228 | 0.866 | -15.6 | -25.1 | -24[28] |
| Zn | 2.867 | 1.217 | -29.5 | -36.6 | 30.5+0.5[68, 69] |
| Mo | 2.528 | 1.377 | -47.3 | -45.9 | |
| W | 2.450 | 1.140 | -41.6 | -45.1 | |
| Sn | 4.007 | 0.176 | -14.2 | -25.1 | |
| Pb | 4.977 | 0.782 | -9.4 | -25.6 | |
| M-HKUST-1 | | | | | |
| Be | 1.945 | 1.367 | -41.3 | -47.5 | |
| Mg | 2.221 | 1.574 | -47.0 | -56.1 | |
| Ca | 2.624 | 1.527 | -50.1 | -51.3 | |
| Sr | 2.861 | 1.482 | -45.9 | -45.1 | |
| Sc | 2.096 | 1.270 | -43.6 | -44.8 | |
| Ti | 2.686 | 1.285 | -30.9 | -30.6 | |
| V | 2.772 | 1.014 | -23.4 | -16.4 | |
| Cr | 3.149 | 1.232 | -20.4 | -22.1 | -26.7[35] |
| Mn | 3.106 | 1.036 | -20.9 | -26.4 | |
| Fe | 3.282 | 0.992 | -12.0 | -26.5 | |
| Co | 2.584 | 1.149 | -22.2 | -28.9 | |
| Ni | 2.731 | 1.041 | -23.6 | -32.3 | -36.8[35] |
| Cu | 2.769 | 0.940 | -17.8 | -30.5 | -23.7±8.2[35, 70-72] |
| Zn | 2.384 | 1.236 | -34.4 | -36.6 | |
| Mo | 3.340 | 1.237 | -17.5 | -26.4 | -25.6[35] |
| W | 3.127 | 1.270 | -15.9 | -19.4 | |
| Sn | 3.811 | 0.270 | -14.1 | -21.7 | |
| Pb | 3.840 | 0.591 | -5.1 | -9.3 | |

**Table 1**. Calculated adsorption enthalpies (kJ/mol $CO_2$), metal-oxygen bond lengths (Å) (involving the nearest oxygen atom in $CO_2$), and metal charge (REPEAT method[62]) for metal substituted-variants of DOBDC and HKUST-1.

In contrast to Be, the other three metals – Cr, Cu, and Sn – exhibit geometries in which the CUS protrudes from the framework, resulting in a change from square-pyramidal-like coordination to square-planar-like with four metal-oxygen bonds. We note that the synthesis and $CO_2$ uptake of Cu-DOBDC was reported during the review of the present manuscript.[28] Consistent with our predictions, in the as-synthesized structure Cu adopts a 4-fold coordination, while the measured adsorption enthalpy (24 kJ/mol) is in excellent agreement with our calculations (25.1 kJ/mol).

Structures for the relaxed SBUs in M-HKUST-1 generally exhibit less distortion than in M-DOBDC. In all cases metal substitution preserves the paddle-wheel geometry (Fig. S3), with the main difference between variants being the position of the metal with respect to the 4-fold oxygen plane. The largest distortions occur for Ca, Sr, Sn and Pb, which we classify as yellow. Metal substitution in these cases results in a large protrusion of the metal out of the oxygen plane, accompanied by an enlargement of the M-M distance (Table S3). The protrusion appears to arise from a size effect related to the large ionic radii of these metals (Table. S4). No M-HKUST-1 structures are classified as red.

Considering the smaller number of red structures in the M-HKUST-1 series, it appears that this compound is more amenable to isostructural metal substitution than the M-DOBDC series. In general, the bonds between the CUS and nearest neighbor framework oxygen are slightly smaller in HKUST-1 than in DOBDC; this trend is consistent with the higher coordination of metal sites in DOBDC (5-coordinated, square pyramidal) vs. HKUST-1 (4-coordinated, square planar).

Table S5 summarizes the structural properties of the $CO_2$-adsorbed state for both MOFs. In the case of M-DOBDC, $CO_2$ adsorption does not induce significant changes to the structure of the MOF regardless of the identity of the CUS metal. Distances between the metal and the nearest oxygen in $CO_2$ vary from a minimum of 2.34 Å in V-DOBDC to a maximum of 4.07 Å in Sn-DOBDC, with shorter bond lengths correlating with larger adsorption enthalpies (Table 1), a trend which has also been observed by others.[24] Adsorbed $CO_2$ exhibits a slight lengthening of the C-O bond closest to the metal in all versions of M-DOBDC, while the distal C-O bond slightly shrinks. (The revPBE-vdW C-O bond length in an isolated $CO_2$ molecule is 1.179 Å.) A small (0-3°) deviation from the linear O-C-O bond angle is also observed.

The existence of four symmetry-equivalent adsorption geometries per metal site in M-HKUST-1[21] results in the possibility for shearing of the paddle-wheel SBU upon adsorption of $CO_2$ (see Fig. 1c for an example involving Mg-HKUST-1). This effect is most pronounced when the two adsorbed $CO_2$ molecules bonded to an given SBU are oriented in a *trans* configuration, and for cases where the metal-$CO_2$ interaction is strongest (i.e., Mg, Ca, Sr and Sc, see below). Shearing could be expected if adsorbed $CO_2$ cannot easily hop between the four sites. We find this is indeed the case, as a rotational energy barrier ($E_a > k_BT$) arising from steric hindrance with the nearby linker prevents easy transitions between adsorption sites. The contribution of shearing relaxations to the $CO_2$ adsorption energy was estimated by comparing to simulations in which these relaxations were forbidden. We find that shearing can lower binding energies (i.e., more exothermic) by up to 3-6 kJ/mol for M = Ca and Sr.

Regarding the geometry of metal-$CO_2$ interactions, M-O bond distances in M-HKUST-1 are generally smaller than in M-DOBDC, ranging from 1.95 Å in Be-HKUST-1 to 3.84 Å in Pb-HKUST-1. Adsorbed $CO_2$ generally exhibits a more linear geometry in M-HKUST-1, suggestive of slightly weaker bonding interactions with the MOF; other features of the $CO_2$ structure follow trends similar to those observed in M-DOBDC. (An exception is Sc-HKUST-1, in which the bending angle of $CO_2$ is 140.8°.)

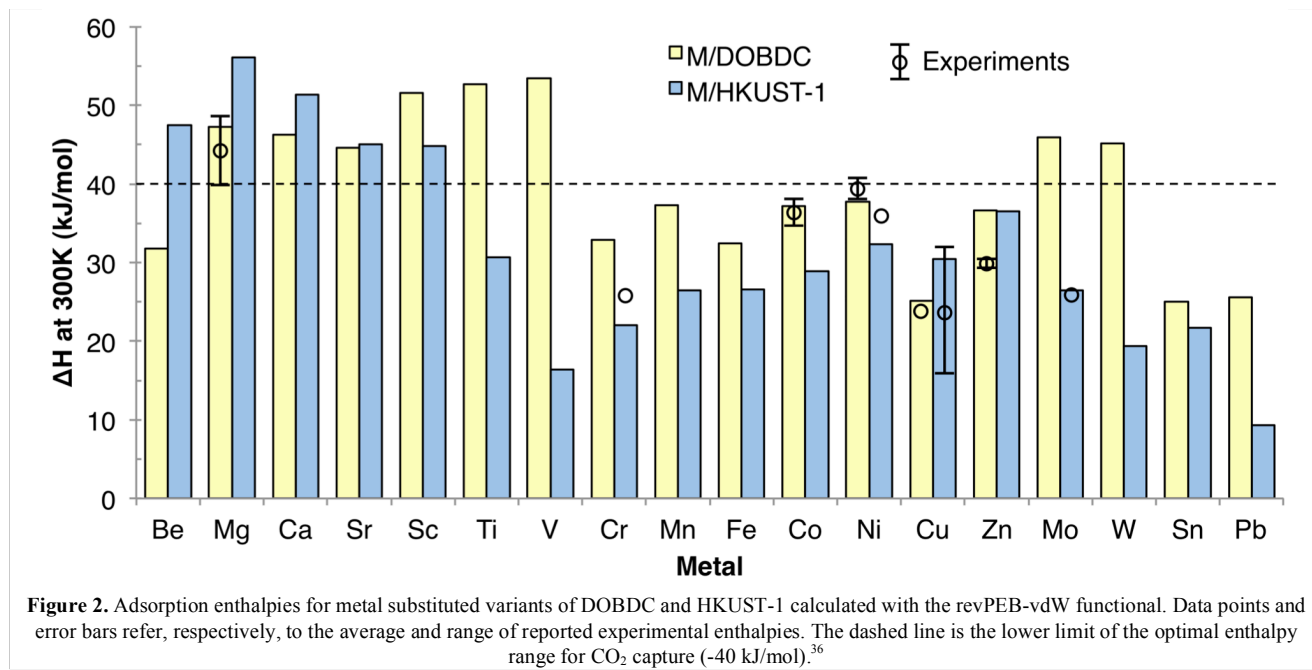

**Figure 2.** Adsorption enthalpies for metal substituted variants of DOBDC and HKUST-1 calculated with the revPEB-vdW functional. Data points and error bars refer, respectively, to the average and range of reported experimental enthalpies. The dashed line is the lower limit of the optimal enthalpy range for $CO_2$ capture (-40 kJ/mol).[36]

As previously described, the metal ions in the relaxed Cr, Cu, and Sn-based DOBDC variants adopt distorted square-planar-like geometries, and it is natural to ask whether these compounds locally resemble the analogous HKUST-1 geometries and possess similar binding energies for $CO_2$. Indeed, in both Cu/Sn-DOBDC and Cu/Sn-HKUST-1 comparable binding energies (Table S6) are found (see Figure S2, S3 and Table S2, S3). While in the case of Cr-DOBDC and Cr-HKUST-1 both compounds exhibit similar square-planar geometries and M-O distances, however, the slight submersion of Cr below the square pyramid plane in Cr-HKUST-1 makes it less accessible and results in weaker binding with $CO_2$.

**Thermodynamics**

Calculated enthalpies for $CO_2$ adsorption at 300 K are tabulated in Table 1 and summarized graphically in Fig. 2. A complete compilation of thermodynamic data (calculated using the LDA[51], PBE-GGA[52], semi-empirical DFT-D2[53] and revPBE-vdW[39]) across all 36 metal substituted MOFs is given in Table S6. For comparison, experimental adsorption enthalpies from the literature are also included in Table 1 and Fig. 2. In the case of Mg, Ni, Co, Zn-DOBDC and Cu-HKUST-1, several experimental measurements have been performed by different groups yielding a relatively robust estimate of thermodynamic properties of these compounds. Experimental data has also recently been reported for Cu-DOBDC, Cr, Ni and Mo-HKUST-1, however only one measurement has been performed for each of these cases. Consistent with prior studies[48], we find LDA and GGA yield rather poor agreement with experimental adsorption enthalpies: the omission of vdW interactions in these methods results in significant under- (GGA) and over-estimation (LDA) of the experimental adsorption enthalpies.

The inclusion of dispersion interactions in DFT-D2 and the revPBE-vdW functionals significantly improves the accuracy of the predicted binding energies. In the case of ΔH, the revPBE-vdW exhibits better agreement with experimental data than the DFT-D2 functional: The average error for revPBE-vdW compared to experiments, 3.4 kJ/mol, is less half that for DFT-D2, 7.3kJ/mol, consistent with our prior benchmarking of these methods.[48] Trends in binding energies across the metals within each MOF prototype are largely captured in each of the PBE-GGA, DFT-D2, and revPBE-vdW functionals; this implies that the degree of dispersion interaction is proportional to the respective PBE-GGA binding energy.

According to a recent report,[38] optimal adsorbents for post-combustion or direct-air carbon capture will exhibit adsorption enthalpies between 40 – 75 kJ/mol. While the targeted range of ΔH is based on considerations related to regeneration efficiency, other authors have suggested[6] that $CO_2$ capacity may also scale with ΔH, suggesting that enthalpies towards the higher end of this range may be desirable. Excluding compounds having large structural distortions (i.e., "red" compounds), inspection of the calculated enthalpies in Table 1 reveals that eight variants of M-DOBDC (M = Mg, Ca, Sr , Sc, Ti, V, Mo, and W) and five variants of M-HKUST-1 (M = Be, Mg, Ca, Sr, and Sc) fall within the targeted thermodynamic window. Of these, [Sr, Mo, W]-DOBDC and all five M-HKUST-1 variants have to our knowledge not been previously identified as promising materials; they therefore represent targets of interest for experimental synthesis and testing. More generally, with the exception of Be-DOBDC (for reasons previously described), substitutions involving alkaline earth metals show promise. While other properties of these materials will certainly be important in assessing their viability in carbon capture applications (e.g., cost, selectivity, robustness to water vapor and other reactive flue gas species, *etc.*), the efficient computational identification of those compounds which hold promise from those which are "thermodynamic dead ends" is clearly of value.

Our predicted energetics qualitatively agree with those of Park *et al*.[24] who calculated the static binding energy (ΔE) for $CO_2$ on a subset of M-DOBDC compounds using the semi-empirical

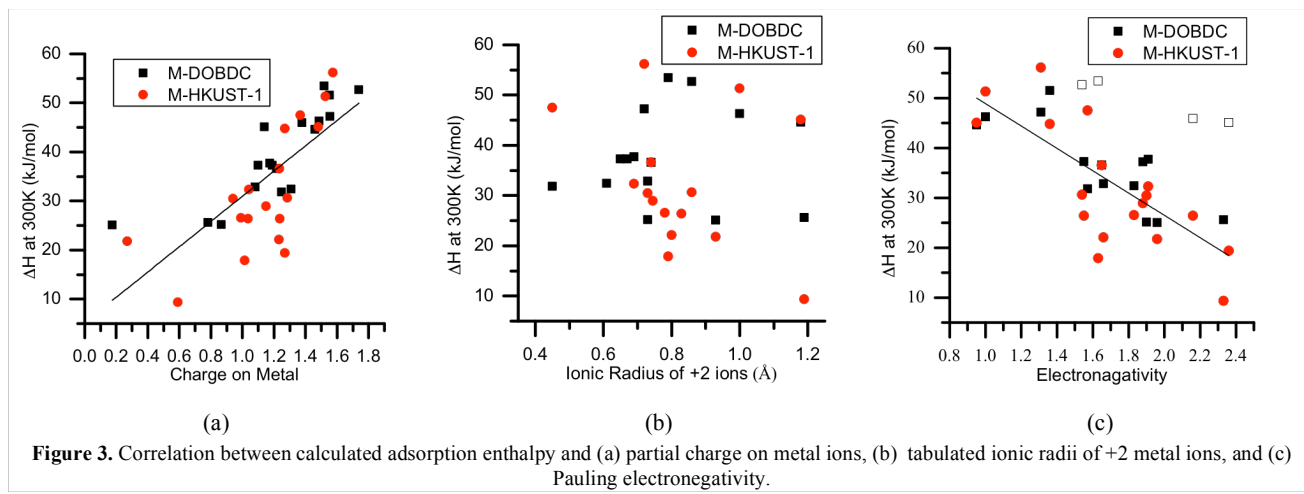

**Figure 3.** Correlation between calculated adsorption enthalpy and (a) partial charge on metal ions, (b) tabulated ionic radii of +2 metal ions, and (c) Pauling electronegativity.

DFT-D2 method. The trends predicted by both methods are similar; nevertheless, the agreement with experimental data is significantly better with vdW-DF as DFT-D2 tends to systematically under-predict binding energies, a feature which was observed in our prior study.[48] The good agreement with experimental data combined with the modest computational cost – vdW-DF calculations are only ~50% more expensive than a conventional GGA calculation – suggests that the vdW-DF method is well suited for efficient characterization of $CO_2$ capture in MOFs.

Incidentally, those MOFs that were identified as less likely to exhibit isomorphism exhibit amongst the smallest $CO_2$ adsorption enthalpies. For example, in M-DOBDC Sn and Cu exhibit enthalpies of only ~25 kJ/mol, and are followed closely by Pb and Be. Similarly, in M-HKUST-1, Pb and Sn fall near the bottom of the range of calculated enthalpies (9.3 and 21.7 kJ/mol, respectively).

**Electronic Structure**

Figure 1 compares the local geometry near the CUS metal (after $CO_2$ adsorption), charge density difference, and local density of states (LDOS) for four MOFs whose behavior spans the range of properties observed in the screened compounds. For M-DOBDC we illustrate the cases M = Mg and Fe. Both of these compounds have been synthesized,[12, 26] but the performance of only Mg-DODBC has been tested with regard to $CO_2$ uptake. Mg-DOBDC exhibits a strong $CO_2$ affinity, -47.2 kJ/mol (Table 1), whereas the interaction between $CO_2$ and Fe-DOBDC falls on the weaker end of the scale, -32.4 kJ/mol. These thermodynamic trends are reflected in the degree of charge redistribution in these compounds (Fig. 1, middle panels): in Mg-DOBDC there is a large accumulation of charge (+0.011 $e$-/Å$^3$) on the $CO_2$ oxygen closest to the CUS metal. This is accompanied by charge depletion on the C and O (in $CO_2$) farthest from the CUS. These features confirm that the Mg CUS in DOBDC induces a large polarization of the $CO_2$ molecule, consistent with the large calculated adsorption enthalpy. This behavior should be contrasted with the charge density of the more weakly bound Fe-DOBDC, which shows a much smaller polarization of $CO_2$ and a maximum charge accumulation (0.004 $e$-/Å$^3$) that is nearly three times smaller than that for Mg-DOBDC.

These data are consistent with previous studies[21, 24] that have argued that strong electrostatic interactions between CUS cations and the $CO_2$ quadrupole constitute the main MOF/$CO_2$ interaction. These interactions result in polarization and a slight bending of the $CO_2$ molecule. In addition, Park et al. also suggested a forward donation of loan-pair electrons in $CO_2$ to metal occurs in M-DOBDC compounds with M = Ti, V. Such an interaction would strengthen the $CO_2$ attraction at these metal sites, resulting in higher adsorption energies.

Turning to M-HKUST-1, Fig. 1 compares the cases having M = Mg and Cu. Mg-HKUST-1 is a hypothetical MOF that our calculations predict as having the largest $CO_2$ adsorption enthalpy within the M-HKUST-1 series, -56.1 kJ/mol (Table 1). On the other hand Cu-HKUST-1 is the well-known prototype for the HKUST-1 series; it has a moderate adsorption enthalpy of -30.5 kJ/mol. Similar to the two DOBDC cases described above, trends in charge density difference distributions for Mg and Cu-HKUST-1 largely follow the calculated binding energies. Adsorption on Mg-HKUST-1 results in a much stronger polarization of the $CO_2$ molecule, with a maximum charge density accumulation (0.016 $e$-/Å$^3$) on the nearest O atom that is more than five times greater than that observed in Cu-HKUST-1 (0.003 $e$-/Å$^3$).

Figure 1 (bottom panel) compares changes to the local density of states (LDOS) of both the CUS metals and $CO_2$ molecules before and after adsorption. In all cases there is no significant change in metal DOS below Fermi level upon $CO_2$ adsorption. In contrast, the $CO_2$ states are uniformly shifted to lower energies upon adsorption. The magnitude of this shift follows closely the trends in adsorption energy, and is further supports the hypothesis that electrostatic effects constitute the primary bonding interaction. These results are in good agreement with the DFT-D2 data of Park et al.[24] Charge density difference and local density of states plots for all compounds examined in this study are provided in Figs. S4 – S9 in the Supporting Information.

In contrast to substitutions involving alkaline earths, where ΔH is relatively smaller for the M-DOBDC-based compounds, Fig. 2 shows that M-DOBDC compounds containing transition metals generally have adsorption enthalpies that are slightly larger than their corresponding M-HKUST-1 variants. Exceptional cases occur for M = Ti, V, Mo, and W, where the difference in ΔH between DOBDC and HKUST-1-based compounds is especially large, exceeding ~ 20 kJ/mol. These differences in affinity are also evident in the charge density difference plots (Figs. S4 and S5): there is significant accumulation of charge between the metal atom and $CO_2$ in the DOBDC-based compounds, whereas in HKUST-1 the accumulation is much smaller. The larger $CO_2$ affinity and charge accumulation associated with these four metals in DOBDC appears to arise from forward donation of electrons from the $CO_2$ HOMO (lone pair electrons) to partially occupied $d$ states on the metal. The reason this donation is more facile in DOBDC than in HKUST-1 can be understood from the local density of states on the metal sites (Fig. S10). In DOBDC the square-pyramidal coordination of the metals allows in some cases for significant state density near the Fermi level. The effect is most significant for Ti, V, Mo, and W; hence these states would be energetically well-suited to accept donated electrons. In contrast, in HKUST-1 the LDOS for essentially all metals is negligible at the Fermi level. Finally, the increase in negative charge on Ti, V, Mo, and W following adsorption, and the simultaneous increase in positive charge on $CO_2$, (Table S7) further suggests that these energetic differences can be traced to the efficacy of forward electron donation.

**Trends**

Figure 2 shows that in two thirds of the possible compounds DOBDC-based structures have a higher affinity for $CO_2$ than those based on HKUST-1. A notable exception to this trend are the variants based on the alkaline earth metals (AEM). These trends can largely be explained by the accessibility and charge state of the CUS metal, as described below.

Figure 3 examines the correlation between adsorption enthalpy and the CUS metal's calculated oxidation state (a), ionic radius (b, Table S4), and electronegativity. Values for the charges are summarized in Table 1. Larger charges on the metal are expected to enhance the electrostatic interaction between MOF and $CO_2$, resulting in higher adsorption enthalpies. We find that the average metal partial charge in DOBDC of +1.22 is larger than the average of +1.14 in HKUST-1, in agreement with the stronger affinities generally observed in DOBDC-based compounds. As demonstrated in the figure 3(a), the magnitude of the metal charge correlates with the adsorption enthalpy, as could be expected given that electrostatic interactions are a significant portion of the MOF-$CO_2$ interaction. In particular, compounds having CUS partial charges larger than approximately 1.4 exhibit adsorption enthalpies within the desired window of ΔH > 45 kJ/mol.[38] Such a correlation suggests that by calculating partial charges alone one could quickly identify MOFs with promising thermodynamics. Such an approach would represent a significant savings over full-scale adsorption calculations that require optimizing adsorption geometries.

In the case of alkaline earth metal (AEM) substitutions, M-HKUST-1 variants are predicted to have larger ΔH than the corresponding M-DOBDC variants. This behavior differs from the general trend mentioned above; nevertheless it still can be attributed to the magnitude of the charges on CUS cations, which are larger in M-HKUST-1 than in M-DOBDC in the case of AEM (Table 1). The absence of $d$-electrons in the AEM allows for relatively higher charge on the CUS in the four-fold-coordination environment of HKUST-1 vs. the five-fold-coordination of M-DOBDC. Additional attraction for $CO_2$ in AEM-HKUST-1 appears to arise from interactions with other atoms in the SBU, as suggested by the larger non-linearity of the O-C-O angle (Table S5).

In addition to the charge on the metal, it has also been suggested[27] that metal ions having small ionic radii should strongly polarize guest molecules, resulting in a more exothermic adsorption enthalpy. Since ionic radii are tabulated quantities (Table S4.), a correlation between radii and $CO_2$ affinity could be easily exploited to direct synthesis efforts towards specific metals. Figure 3(b) plots calculated adsorption enthalpies vs. the tabulated ionic radius of the 2+ ions used in this study. The figure demonstrates that in these compounds there is no general correlation between ΔH and ionic radius. This behavior can be understood by noting that ionic radius will also impact the structure of the SBU. A good example is Be. As the element having the smallest ionic radius, one would expect that Be will have a strong polarization effect on $CO_2$, resulting in a large ΔH. While the expected behavior holds for Be-HKUST-1 (ΔH = -47.5 kJ/mol), Be-DOBDC has a low enthalpy of ΔH = -31.8 kJ/mol. This is due to size effects within the SBU: the small radius of Be ions results in a "burrowing" of the Be ion into the framework, where it adopts a tetrahedral coordination and is no longer readily accessible to adsorbed $CO_2$. While correlations between ΔH and radius could be expected across some subset of the examined compounds where the structure effects are small (for example, the trend holds for Mg, Ca, and Sr), in general the connection between $CO_2$ affinity and ionic radius of the CUS is complicated by changes to the MOF structure arising from ion size effects.

As the electrostatic interaction between the MOF's metal cations and adsorbed $CO_2$ molecules appears to comprise a significant fraction of the adsorption enthalpy, it is natural to ask whether the electronegativity of the metal correlates with ΔH. Presumably, those metals that are the least electronegative will exhibit the largest positive partial charges due to ionic interactions with the MOF ligands; in turn, the induced metal charge should result in stronger interactions with $CO_2$. Figure 3(c) plots the tabulated Pauling electronegativities of the metals vs. the calculated ΔH values. With the exception of Ti, V, Mo, and W-DOBDC, for which forward donation constitutes a significant fraction of the bonding (shown as empty symbols in the figure), it is clear that a correlation exists between enthalpy and electronegativity. Such a correlation suggests that MOFs containing electropositive metal ions will have amongst the highest affinities for $CO_2$.

**Conclusion**

Van der Waals-augmented DFT has been used to screen 36 metal substituted variants of M-DOBDC and M-HKUST-1 (M = Be, Mg, Ca, Sr, Sc, Ti, V, Cr, Mn, Fe, Co, Ni, Cu, Zn, Mo, W, Sn, and Pb) with respect to their $CO_2$ adsorption enthalpy at T = 300K. The prototype compounds were selected based on their high capacities for $CO_2$, their potential for forming isostructural metal-substituted variants, and to examine the impact of their distinct metal cluster geometries (square-planar vs. square pyramidal). An analysis of the structure of the metal cluster was used to qualitatively assess the likelihood that a given substituted metal will adopt an isostructural geometry, and suggests that the M-HKUST-1 structure is more amenable to metal substitution than is M-DOBDC.

Consistent with our prior benchmarking, enthalpies calculated with the non-empirical revPBE-vdW functional are in good agreement with experimental measurements, and suggest that this functional is a reliable and efficient method for treating the large unit cells typical of MOFs. Electronic structure trends across the metals reveal that electrostatic interactions comprise a significant portion of the MOF-$CO_2$ bond, in agreement with several literature reports. These trends further suggest that the geometric accessibility and partial charge of the CUS metal correlates with the magnitude of the adsorption enthalpy. *Thus, the metal charge could be used as a simple descriptor to rapidly identify MOFs with targeted adsorption enthalpies without the need for expensive adsorption calculations.* The dependence on the metal's charge state is further reflected in a correlation with the metal's electronegativity, suggesting that strongest affinities will be obtained for MOFs containing the most electropositive metals. On the other hand, due to structural effects the ionic radius of the CUS metal does not generally correlate with the adsorption enthalpy: extremely small and large ions alter the structure of the MOF, potentially limiting the accessibility of these ions to adsorbed $CO_2$.

Finally, our calculations identify several compounds having $CO_2$ affinities that fall within the targeted range of -40 to -75 kJ/mol. While other properties of the identified compounds need to be assessed (stability, selectivity, *etc*.), the ability to rapidly distinguish promising compounds from those that are "thermodynamic dead-ends" via computation will be of value in guiding synthesis efforts towards promising compounds.

## Notes and references

*[a] Mechanical Engineering Department, University of Michigan, Ann Arbor, Michigan 48109, United States. E-mail: djsiege@umich.edu*

† Electronic Supplementary Information (ESI) available: Calculated binding energies and detailed geometric (ball and stick models of local structure near the metal site) and electronic structure information (charge density difference plots, and local DOS plots) for all compounds considered.